An ALE meta-analytic comparison of verbal working memory tasks

Timothy J. Wanger
University of Georgia

Keywords: fMRI, meta-analysis, 2-back, 3-back, PASAT

ABSTRACT

**Background**: The n-back and Paced Auditory Serial Addition Test (PASAT) are commonly used verbal working memory tasks that have partially overlapping uses in clinical and experimental psychology. We performed three activation likelihood estimation (ALE) meta-analyses, comparing two load levels of the n-back task (2-back, 3-back) to the PASAT and to each-other. These analyses shed light on the recruitment of verbal working memory regions, as well as overall involvement of cognitive and emotional brain regions in these tasks.

**Methods:** ALE meta-analysis was performed on published studies that 1) used healthy adults, 2) had >5 subjects, 3) used whole-brain fMRI neuroimaging and 4) reported coordinates for the 2-back, 3-back, or PASAT.

**Results and Implications**: The results from analyses of 68 verbal working memory studies revealed higher overall likelihood of activation the frontal eye fields in the 3-back. The PASAT exhibited higher overall activation in the bilateral supplementary motor areas (SMA), left supramarginal gyrus, and left superior parietal lobule. Furthermore, the 3-back exhibited higher activation in the right SMA, and anterior mid-cingulate cortex versus the 2-back, and the PASAT exhibited higher activation in a cluster near the right premotor area versus the 2-back. A laterality effect was observed in the dorsolateral prefrontal cortex between the PASAT (left) and 3-back(right). These data suggest greater activation of regions traditionally associated with the phonological loop during the PASAT, compared to the 2- and 3-back tasks. Furthermore, individual ALE analyses suggest involvement of emotional processing and salience network regions (insula, cingulate) in addition to the well-established verbal working memory regions (Broca's region, bilateral SMA, premotor, posterior parietal cortices) in all 3 tasks. Lateralized activation in dorsolateral prefrontal cortex contributes to a growing literature of hemispheric effects in this particular region.

**Limitations:** The methodological shortfalls of ALE meta-analyses lead to an unavoidable degree of spatial uncertainty for the results. Power could be improved in future iterations with the proliferation of more studies using these tasks. However, these findings are in close agreement with prior work (Owen 2005), inspiring a reasonable degree of confidence.

**Conclusions:** These meta-analyses provide the first glimpse into the regions activated by the PASAT, which has not been meta-analytically reviewed prior to this study. Further, these data illustrate the sensitivity of ALE meta-analysis to identify differences in activation across verbal working memory studies that may be associated with specific cognitive and emotional aspects of these tasks. Further parametric work examining these tasks is necessary to determine more precisely the causes of the activation patterns revealed here.

## 1. Introduction

Verbal working memory (VWM) is broadly defined as the set of cognitive processes responsible for manipulating and storing recently perceived phonological or acoustic information in short term memory. This information can be used immediately to inform an appropriate action or be transferred to long term memory (Baddeley 2003). In the most prominent model of working memory, VWM is characterized as a "phonological loop", a brain system subordinate to a central executive network (Baddeley 1974). The phonological loop is composed of two functional systems, a phonological store and articulatory rehearsal network. These components are thought to form a metaphorical 'blackboard of the mind,' which couples auditory perception with cognitive processing and action (Baddeley 2003, Buchsbaum 2008). Many tasks have been designed to test features and limitations of verbal working memory in different ways, and the emergence of neuroimaging techniques has quickly identified the basic neural correlates of VWM. However, more parametric work is needed to understand the intricate differences between VWM tasks.

Two prominent tasks used to study VWM are the n-back and Paced Auditory Serial Addition Test (PASAT). These tasks have been used to assess VWM function in clinical and healthy populations. Each task requires that participants hold multiple phonological chunks of information in working memory, manipulate the information, and relay an answer to the experimenter (Gronwall 1974, Tombaugh 2006). The tasks differ in four key respects: 1) The n-back uses a sequence of letters, while the PASAT uses a sequence of numbers. 2) The n-back presents information visually, while the PASAT predominantly presents information auditorily. 3)

The manipulation is different between the two tasks. In the n-back, participants must remember if the current letter is the same as the letter 'n'-times back. In contrast, participants must add the two most recent numbers during the PASAT. 4) Participants respond via keypress for the n-back and respond auditorily for the PASAT. Moreover, the n-back is designed to have different levels of difficulty (i.e. 2-back, 3-back), where the PASAT is not. These task differences span multiple dimensions and as a result may engage cognitive, motor, and affective brain regions in varying degrees.

Both tasks have been used to identify association to VWM deficiencies in specific brain disorders. The PASAT is commonly used as a neuropsychiatric assessment tool, where poor performance is an indicator of Multiple Sclerosis (MS; Gontkovsky 2006, Rosti 2007) and the duration of persistence is related to functional brain connectivity, abstinence, and cessation in substance users (Brown 2002, Daughters 2005, Daughters 2017). Similarly, poor n-back performance is seen in MS patients and traumatic brain injury (TBI; Parmenter 2006, Perlstein 2004, McAllister 2001), and is linked to altered brain activation in MS patients and smokers (Sweet 2004, Sweet 2006, Sweet 2010). These tasks are thought to measure the psychological construct of processing speed (Parmenter 2006, Redick 2013), which may be worse due to axonal injury in MS and TBI (Rao 1986, Rao 2000).

However, participants report that the PASAT is more stressful than the n-back (Parmenter 2006), and it has been used to induce negative affective states (Deary 1994, Holdwick 1999). In smokers, rather than a brain processing speed deficiency, poor performance is thought to be related to cognitive and emotional ability to tolerate distress (Leyro 2010,

Zvolensky 2011). In fact, a strongly supported theory of nicotine dependence suggests that the inability to tolerate the cognitive and emotional distress associated with withdrawal is a key factor in relapses (Brown 2005). While the 2-back is also reportedly stressful, it is not as stressful as the PASAT (Parmenter 2006) and does not induce negative affect as measured by the Positive and Negative Affect Schedule (PANAS; Watson 1988, Scott 2015). This affective response has the potential to confound the results of the verbal working memory task - being distressed has been shown to reduce behavioral performance on working memory tasks (Schoofs 2008). Thus, it is important from an experimental validity standpoint to examine the cognitive and affective influences present in these tasks.

Previous meta-analyses have reported significant effects of task design, cognitive load, and stimulus modality on brain activity between the n-back and other working memory tasks (Owen 2005, Rottschy 2012), but no existing meta-analyses have examined the PASAT. It is unclear which emotion-related brain regions might be active during the PASAT, or how the PASAT compares to the n-back in terms of cognitive load. The functional correlates of working memory have been studied during various cognitive load levels of the n-back (1-back, 2-back, 3-back). Some have reported an inverted U-shaped curve of activity in left dorsolateral prefrontal cortex with increasing load (DLPFC; Callicott 1999), while others have reported a linear increase in bilateral DLPFC activity with load (Pochon 2002, Yun 2010). While the parameters of the PASAT are most similar to a 2-back, it is unclear whether brain activity during the PASAT would be more similar to the 2-back, or 3-back. Importantly, performance on the n-back gets progressively worse with increases in difficulty (Callicott 1999, Jaeggi 2010). Several theories regarding performance have suggested that error processing and conflict monitoring activate part of the anterior cingulate, a core region of the emotional processing and salience networks (Botvinick 2001, Holroyd 2002, Etkin 2011).

Another previously mentioned cognitive component of the PASAT that is absent in the n-back is mathematical manipulation, which engages both cognitive and affective resources. Many studies have used mental arithmetic tasks such as the "Montreal imaging stress task" and "Trier social stress task" as a means of inducing stress. Arithmetic tasks have been linked to functional activation in the angular gyrus (Dedovic 2005, Pruessner 2008), in addition to the broad fronto-parietal working memory network. This is the case in studies using the PASAT, which find robust activation of the angular gyrus (Audoin 2005, Lockwood 2004), but not meta-analytic studies of the n-back (Owen 2005, Rottschy 2012). The precise nature of how mathematical manipulations activate affective regions is not well understood.

Several emotion-related regions have been linked to working memory tasks, but more research is needed to fully understand neural mechanisms involved. For instance, a majority of the studies using the PASAT identify activation in the anterior cingulate, but there is uncertainty as to whether this is reflective of task difficulty, cognitive conflict, inhibition of a prepotent response, or other function (Audoin 2005, Mainero 2004, Smith & Jonides 1997). The anterior insula, which is known to be recruited during emotional perception and while performing emotionally demanding cognitive tasks (Bush 2000), is reliably activated during the n-back. However, the activation of this region is inconsistant for the PASAT (Rottschy 2012, Owen 2005, Audoin 2005, Cardinal 2008, Archbold 2009). Additionally, the amygdala is active during emotional processing (LeDoux 2003) and is thought to be a target for emotional regulation during cognitive tasks (Goldin 2008, Davidson 2000). Specifically, the amygdala has reciprocal projections with a network of frontal regions including the dorsomedial prefrontal cortex, ventromedial prefrontal cortex, cingulate, and orbitofrontal cortex (Banks 2007). The amygdala shows a reduction in activity during high-load instances of the n-back (Pochon 2002, Yun 2010), but there are few reports of amygdala activation during the PASAT. It is possible that amygdala activity is attenuated as a result of prefrontal regulation (Banks 2007), but perhaps this regulation is not as successful as during the n-back, since the PASAT reportedly induces a negative mood. In order to compare the relative cognitive and affective regional activation between the n-back and PASAT, we performed Activation Likelihood Estimation (ALE) meta-analyses. In order to test the effect of increasing cognitive load, we contrasted two versions of the n-back (the 2-back and 3-back) with the PASAT and with each other.

Prior functional neuroimaging studies have generated a strong foundational understanding of VWM and its components, which provided us with basic expectations for what we would see in individual task ALEs. For instance, the phonological store is putatively localized in the posterior parietal cortex (Paulesu 1993, Awh 1996, Smith 1998, Buchsbaum 2008). Several Brodmann's areas (BA) in this region have been linked to distinct VWM functions. The superior parietal lobule (BA 7; Becker 1999, Awh 1996) is thought to be involved in visual attention for tasks that have rapidly changing visual elements, while the supramarginal gyrus is more closely related to phonological processing (BA 40; Paulesu 1993). VWM tasks also activate a network of frontal regions that includes the supplementary motor area (SMA), Broca's regions, prefrontal, premotor, cingulate, as well as the cerebellum (Cohen 1997, Callicott 1999, Sweet 2004, Sweet 2006). The cerebellar, premotor, SMA, and Broca's regions are thought to mediate auditory rehearsal and speech (Awh 1996, Smith & Jonides 1997), forming the aforementioned articulatory rehearsal network.  Additionally, regions of the prefrontal cortex are well known for their involvement in executive functions, such as attention and manipulation of information (Baddeley 2003, D'esposito 1995).  Together, these regions form a core network for the n-back identified by a prior meta-analyses (Owen 2005).

Our *a-priori* hypotheses were that we would see common activation in core verbal working memory regions for each of the three tasks. This includes the supplementary motor area, Broca's regions, bilateral prefrontal, bilateral premotor, bilateral cingulate, and bilateral parietal cortices. For the PASAT, we predicted task-specific regional activation in the angular gyrus, amygdala, and anterior cingulate cortex.  When comparing the 3-back to the 2-back, we hypothesized that there would be elevated activation in some regions of the fronto-parietal working memory network.  From prior literature, we expected to see elevated dorsolateral prefrontal cortex activation and decreased amygdala activation due to higher cognitive load (Smith & Jonides 1997, Calicott 1999, Yun 2010).  When comparing the PASAT to the 2- and 3-back, we expected to see greater angular gyrus activation (Audoin 2005, Lockwood 2004). Furthermore, we hypothesized that there may be greater activation of emotion-related regions in the PASAT > 2-back contrast, relative to the PASAT > 3-back contrast, due to the elevated difficulty and related stress of the performing the 3-back.

**2. Methods**

*2.1. Protocol*

The guidelines outlined by the *Preferred Reporting Items for Systematic Reviews and Meta-Analyses (PRISMA)* statement (Moher 2009) were followed for this meta-analytic review. A PRISMA summary of article identification and screening for this study can be found in Supplementary Figure A.  Analysis of coordinate data was performed using GingerALE v2.3.6, using the instructions available in the program's manual, which is readily available at the program's website (http://brainmap.org).

*2.2. Search Strategy*

Articles were identified in November 2017 by entering the following search strings into PubMed's database search engine:

"((((((n-back[All Fields] OR 2-back[All Fields]) OR 3-back[All Fields]) OR 4-back[All Fields]) AND ("magnetic resonance imaging"[MeSH Terms] OR ("magnetic"[All Fields] AND "resonance"[All Fields] AND "imaging"[All Fields]) OR "magnetic resonance imaging"[All Fields] OR "fmri"[All Fields] OR "neuroimaging"[All Fields] OR "activation"[All Fields]) AND (Journal Article[ptyp] AND "humans"[MeSH Terms] AND English[lang] AND "adult"[MeSH Terms])))))"

"(((((((((PASAT[All Fields] OR mPASAT[All Fields] OR mPVSAT[All Fields] OR "PVSAT"[All Fields] OR "paced auditory serial addition task"[All Fields] OR "paced visual serial addition task"[All Fields]) OR "paced auditory serial addition test"[All Fields] OR "paced visual serial addition test"[All Fields] AND ("magnetic resonance imaging"[MeSH Terms] OR ("magnetic"[All Fields] AND "resonance"[All Fields] AND "imaging"[All Fields]) OR "magnetic resonance imaging"[All Fields] OR "fmri"[All Fields] OR "neuroimaging"[All Fields] OR "activation"[All

*Fields]) AND (Journal Article[ptyp] AND "humans"[MeSH Terms] AND English[lang] AND "adult"[MeSH Terms])))))))))"*

Articles were screened to determine if they met eligibility criteria (Section 2.3) and the reference sections for viable studies were reviewed to identify additional candidate studies.

*2.3. Eligibility criteria*

Published journal articles were included if they met the following criteria: 1) participants were healthy adults with an average age between 18 and 65. 2) there were at least 5 subjects included in the analysis. 3) the article was available in English. 4) whole-brain functional magnetic resonance imaging was used. 5) the article contains activation foci coordinates listed in Talairach or MNI space. Studies of clinical populations and of gene polymorphisms within healthy subjects were excluded. However, clinical studies using healthy adults as a control sample were included if coordinates were listed separately for the control group.

*2.4. Baseline contrast exclusion criteria*

Several types of control tasks are used for PASAT and n-back contrasts. The most frequently used control task for the n-back is the 0-back, where participants are asked to assess whether the presented letter is a pre-specified letter (e.g. Is it an 'X'), indicating yes or no via keypress. For the PASAT, participants are most commonly asked to repeat the letter that is presented to them. Studies using the 2-back were excluded if their control tasks were drastically different from the norm, or if the tasks were being performed under additional external demands or conditions. These exclusion criteria were relaxed for the 3-back and PASAT studies, due to the relative scarcity of studies available for these tasks. While this is not optimal, some heterogeneity is expected in task design to begin with, and this choice was motivated by recent work recommending at least 17 studies for enough power to detect small effects (Eickhoff 2016). The 3-back meta-analysis includes two studies using the 1-back as a contrast, which were included for these reasons. While this control task is more demanding than the 0-back, it is considered to be very easy. Performance on the 1-back is nearly as high as the 0-back (~99%), and substantially higher than the 2- or 3-back (Callicott 1999, Jaeggi 2010).

*2.5. ALE meta-analysis process*

Activation likelihood estimation (ALE; Turkeltaub 2002) is a meta-analytic technique that is used to combine and compare results from many studies. This method generates an approximate representation of the results of prior work by utilizing reported coordinates of activation, modeling each foci as spherical Gaussian probability density functions with a full-width half-maximum = 10mm. When foci from a single study are combined, these data form a map where the score for each voxel represents its relative likelihood of activation. Combining the ALE maps from many studies that use the same task generates a map revealing which regions are most consistently activated across the set of studies. This technique can be applied to aggregate data for different tasks or experimental variables, allowing the meta-analytic comparison of factors that were not the primary focus of the prior studies (Swick 2011, Redick 2013). For instance, a meta-analysis of 24 working memory studies using variants of the n-back found differential brain activation when subjects monitor the stimuli's location versus its identity, and when the stimuli itself is verbal or non-verbal (Owen 2005).

An ALE meta-analytic comparison of verbal working memory tasks                                    13In order to determine statistical significance for ALE maps, permutation testing is used (Laird 2005). This generates a null distribution of scores for each voxel, which the experimentally generated ALE score is compared against. To address the problem of multiple comparisons, a cluster-based permutation method is used (Eickhoff 2012). For individual task ALEs, a relatively strict cluster-level family-wise error threshold of 0.05 was used for cluster formation. We then permuted these data 500 times, and used an uncorrected P < 0.001 for statistical inference. These specifications are thought to produce fewer spurious clusters than using a false-discovery-rate correction or uncorrected p-value (Eickhoff 2016). For contrast analyses, our data was permuted 10000 times, using a minimum cluster size of 200mm$^3$. Maps of differential activation between tasks were thresholded by an uncorrected P < 0.01 for statistical inference.

Additionally, regions whose total cluster volume is less than or equal to 40 mm$^3$ are not discussed in the results section. This cutoff represents 20% of the 200mm$^3$ minimum cluster-size threshold used for statistical analysis and was chosen to exclude small portions of clusters that extend into nearby anatomically-defined regions. A full listing of regions comprising each activated cluster and extrema is provided in Supplementary Materials, as generated by the GingerALE program. Too few studies reported deactivation coordinates for each task, making meta-analyses of deactivated regions infeasible. Therefore, all meta-analyses presented herein refer to differences in activated regions.

**3. Results**

*3.1. Article Inclusion*

These meta-analyses included thirty-four 2-back studies (414 foci), seventeen 3-back studies (206 foci), and seventeen PASAT studies (321 foci). Information about each study included can be found in Table 1. All subjects included in these meta-analyses were healthy adults, between the age of 18 and 65. The total subjects included was 1,372 (2-back: n = 885, 3-back: n =266, PASAT: n = 221). The ratio of men to women was larger in the 2-back (491:369) meta-analysis, but nearly equal for the 3-back (116:115) and PASAT (102:104) meta-analyses.

The average ages for each group were compared using a one-way ANOVA and found to be non-significant at the p<.05 level [F(2,62) = 2.92, p = 0.061]. Task performance was also compared using a one-way ANOVA, finding a difference in means [F(2,58) = 4.15, p = 0.021]. A post-hoc Tukey HSD test revealed that the 3-back had worse performance than the 2-back (p = 0.018). Removing the handful of PASAT studies that used simplified designs (which led to 95-100% performance rate) did not change the results of these statistical tests. Study demographics and performance is reported in Supplementary Table 1.

*3.2. Task-specific ALEs*

ALE meta-analysis of the 2-back revealed increased likelihood of activation in bilateral superior frontal gyri, frontal poles, medial frontal gyri, middle frontal gyri, precentral gyri, inferior frontal gyri (including Broca's area), mid-cingulate cortex, claustrum, insulae, inferior parietal lobules, superior parietal lobules, supramarginal gyri, angular gyri, and precunei.

The 3-back had increased likelihood of activation in the bilateral superior frontal gyri, frontal poles, medial frontal gyri, middle frontal gyri, mid-cingulate cortex, inferior parietal lobules, and superior parietal lobules. Lateralized activation was found in the right inferior frontal gyrus, right claustrum, right insula, left angular gyrus, and left precuneus.

The PASAT had increased likelihood of activation in the bilateral superior frontal gyri, medial frontal gyri, precentral gyri, inferior frontal gyri (including Broca's area), superior parietal lobules, and precunei. Lateralized activation was found in the left middle frontal gyrus, right mid-cingulate gyrus, left inferior parietal lobule, left angular gyrus, and left supramarginal gyrus.

ALE activation maps for these tasks are presented in Figure 1. Clusters identified by contrast analyses are described below, from the largest cluster to the smallest cluster within each section. Brodmann areas, cortical anatamy as defined by GingerALE's output, and relevant nomenclature are identified for each cluster. Cluster volume, center of mass coordinates, and regional information is summarized in Table 1 and depicted in Figure 2.

*3.3. 2back > 3back,  2back > PASAT*

No significant clusters of activation were identified for the 2-back > 3-back contrast, or 2-back > PASAT contrast.

*3.4. 3-back > 2-back*

The 3-back exhibited increased activation relative to the 2-back in one cluster that includes the right medial frontal gyrus (BA6), right superior frontal gyrus (BA8), and bilateral mid-cingulate gyrus (BA32).

*3.5. 3-back > PASAT*

The 3-back exhibited increased activation relative to the PASAT in two clusters, including the right middle frontal gyrus (BA9), and right superior frontal gyrus (BA8).

*3.6. PASAT > 2-back*

The PASAT exhibited increased activation relative to the 2-back in four clusters. The first cluster includes the bilateral medial frontal gyrus (BA6), and superior frontal gyri (BA6). The second cluster includes the right precentral gyrus (BA6) and right inferior frontal gyrus (BA9). A third cluster was found in the left inferior parietal lobule (BA40), and the fourth was found in the left superior parietal lobule (BA7).

*3.7. PASAT > 3-back*

The PASAT exhibited increased activation relative to the 3-back in four clusters. The first cluster includes the bilateral medial frontal gyrus (BA6) and left superior frontal gyri (BA6). The second cluster was found in the left middle frontal gyrus (BA46/BA9). The third cluster included the left inferior parietal lobule (BA40), and left supramarginal gyrus (BA40), and the fourth cluster was found in the left superior parietal lobule (BA7).

**4. Discussion**

*4.1. Task-specific ALE maps*

The ALE meta-analyses for the 2-back, 3-back, and PASAT demonstrate a greater likelihood of activation in brain regions commonly associated with working memory tasks as we hypothesized (Figure 1). The activation of bilateral SMA, premotor, frontal, and parietal cortices during the 2-back closely resemble a prior meta-analytic study of working memory (Owen 2005). These findings illustrate the replicability of past ALE meta-analyses using current specifications. Despite significant differences in ALE protocol, study inclusion/exclusion criteria, and having only one study common between the analyses, the similarity of these findings indicate consistent activation of the VWM network. The broad pattern of activation appears to be similar for the three tasks, with only a handful of clusters proving to be significantly different between them.

An additional noteworthy finding from each individual ALE is the degree of involvement of emotion-related regions, such as the insula and cingulate cortex. The high likelihood of activation of these regions in each task indicates that perhaps emotional salience is not entirely separated from the cognitive processes during working memory tasks. Our hypothesis regarding reduced amygdala activation was unable to be tested, since we were unable to perform analyses on deactivated regions. Furthermore, our hypotheses about the PASAT exhibiting elevated activity in the angular gyrus was not supported, since all three tasks displayed activity in the angular gyrus.

*4.2. 3-back > 2-back contrast*

The 3-back had a higher likelihood of activating the right supplementary motor area, right frontal eye field, and bilateral mid-cingulate cortex. The supplementary motor area is functionally heterogeneous, being active in motor, sensory and working memory tasks (Chung 2005). These regions are thought to maintain visuospatial attention during working memory tasks (Owen 2005). Specifically, the frontal eye fields and supplementary motor area work in tandem to hold information 'on-line' and prepare for a motor response (D'Esposito 1998). We interpret this finding as reflective of the greater attention to, and maintenance of information required by the 3-back. This is consistent with previous work showing that the supplementary motor area has increasing activation as a result of load (Smith & Jonides 1997).

Many studies have sought to understand the function of sub-regions of the cingulate cortex (Vogt 2003, Vogt 2004, Etkin 2006, Etkin 2011). The anterior mid-cingulate cortex (aMCC) – often called the rostral cingulate cortex or rostral cingulate zone, has been linked to internally selected actions (Mueller 2007), and the cognitive control responsible for monitoring of response conflict (Barch 2001). Activity in this region is seen during three forms of response conflict: when participants commit errors, when they actively inhibit a prepotent response, and when they choose among a set of equally permissible responses (Botnivick 2001). The set of studies used here had worse behavioral performance on the 3-back, which is potentially responsible for the effect in the aMCC according to the response conflict model.

Early models of the aMCC propose a dichotomy where the dorsal/caudal cingulate is involved in cognitive functions, and the ventral/rostral cingulate is involved in affective processing (Stevens 2011). However, others have suggested that anterior and subgenual cingulate involved in the perception and production of emotion, and experiences of intense negative affect. One model proposes that the anterior and medial cingulate are involved in the appraisal and expression of negative emotion, while the ventral prefrontal regions are involved in regulation of negative affect (Etkins 2011). In our view, this model is not supported by these data, since the PASAT is thought to induce greater negative affect and did not appear to have significantly higher cingulate activity than either the 2-back or 3-back.

*4.3. 3-back > PASAT contrast*

Similar to the 3-back > 2-back contrast (Section 4.2), the 3-back has a higher likelihood of activation in the right frontal eye fields (FEF) when compared to the PASAT. This suggests a greater involvement of visual attention during the 3-back. These results are unlikely to be the result of a simple difference in the mode of stimulus presentation. Visual presentation of VWM stimuli was uniform for all 2-back and 3-back studies included here, while the PASAT sample contained both auditory and visual presentation. If the mode of presentation was driving this effect, we would expect to see greater ALE scores for the 2-back's FEF, compared to the PASAT (which we did not). Prior work investigating stimulus modality effects did not find divergent activation in the FEF. The authors did find an effect in the left posterior parietal cortex during visual VWM presentation and left DLPFC during auditory VWM presentation (Crottaz-Herbette 2003).

      The DLPFC is consistently activated by verbal working memory tasks (Curtis 2003, Sweet 2006). Anterior portions of the DLPFC, located on the middle frontal gyrus, are thought to play a more abstract role in planning, monitoring, and manipulation of working memory information while posterior DLPFC is thought to rapidly adapt to incoming stimuli (Hampshire 2009). For the cluster we identified here, activation in the right anterior DLPFC indicates higher level executive functions are required for the 3-back, compared to the PASAT. With respect to lateralization of activity to the right hemisphere, proponents of the HERA model (Nyberg 1996, Habib 2003) would suggest that this corresponds to episodic memory retrieval, while left lateralized activity occurs during semantic retrieval and episodic memory encoding (discussed in *Section 4.5*). The cluster identified here may be indicative of a cognitive strategy for the 3-back that utilizes visual episodic memory encoding and retrieval. Importantly, lateralized anterior DLPFC activation was

not seen in comparisons to the 2-back, thus this effect is not presumed to be an effect of working memory load.

*4.4. PASAT > 2-back contrast*

The next two contrast analyses indicate that core regions of verbal working memory are more consistently activated during the PASAT when compared to the 2/3-back tasks. Three clusters representing the bilateral SMA, left superior lobule, and left supramarginal gyrus were found in the comparison of the PASAT vs the 2/3-back, and discussion of these clusters applies to both contrasts. Differences were found in the likelihood of activation of prefrontal and inferior frontal regions, which will be discussed independently.

The higher likelihood of activation in bilateral SMA in the PASAT seems to conflict with the results from *Section 4.2*, where elevated SMA activity was seen in the 3-back versus 2-back. However, close examination of these clusters indicates that the cluster referred to here is more posterior than the cluster from *Section 4.2*. Prior work has suggested that rostral SMA is more involved in word generation and working memory tasks, while the caudal aspect is more related to motor and sensory tasks (Chun 2005). Our results appear to conflict with these findings, and more research into this effect is warranted.

Higher likelihood of activation in left parietal regions supports the view that the PASAT engages verbal working memory networks to a greater degree than the n-back. The left superior parietal lobule is known to be active during encoding and rehearsal during verbal working memory (Veltman 2002, Paulesu 1993, Awh 1996, Smith 1998, Buchsbaum 2008). The left inferior parietal lobule is composed of the supramarginal gyrus (BA40) and the angular gyrus (BA39), which are important regions for phonological processing, and mapping sounds onto meaning (Seghier 2013, Burton 2005). The cluster seen here is fully located within the left supramarginal gyrus (BA40), which is thought to be a subsystem for the storage of phonological working memory (Becker 1999, Baldo 2006). Our *a-priori* hypothesis predicted a higher likelihood of activation in the angular gyrus for the PASAT, which is commonly associated with mental arithmetic and calculation tasks (Zago 2002). However, these results indicate that the verbal working memory storage demands are particularly stronger in the PASAT, compared to the 2- and 3-back.

Lastly, the posterior region of the DLPFC, located in the right inferior frontal gyrus (BA9) is thought to be part of a ventral circuit active in attentional orienting, which responds to items that are relevant for immediate behavior (Corbetta & Schulman 2002, Hampshire 2009). This cluster includes both the right inferior frontal gyrus, as well as the somatotopically mapped right premotor cortex (precentral gyrus), which is active during speech production (Pulvermüller 2006). The inferior frontal gyrus is thought to be involved in phonological processing, as part of the articulatory rehearsal network (Burton 2001). While these findings do not highlight Broca's area as being more active during the PASAT, we interpret them as indicating a more active maintenance of phonological representations, which is consistant with the extant literature (Paulesu 1993).

*4.5. PASAT > 3-back contrast*

An ALE meta-analytic comparison of verbal working memory tasks                                    22To reiterate, the higher likelihood of activation in the bilateral SMA, left superior parietal lobule, and left supramarginal gyrus suggest that the PASAT is more consistently activating core verbal working memory regions compared to the n-back task.

However, this meta-analysis identified lateralized activation in the *left* anterior DLPFC. This may be indicative of a different cognitive strategy for performing the PASAT, than the 3-back (see Section *4.3.*). Lateralized activation in left DLPFC is thought to be related to semantic memory retrieval, or episodic memory encoding according to the HERA model (Nyberg 1996, Habib 2003). The PASAT requires participants to respond with declarative information (the sum of two previous numbers), which may explain the higher likelihood of activation in this region. Further work is needed to conclusively determine the causes of the DLPFC asymmetry observed here.

*4.6. Strengths and Limitations*

An advantage this study has over prior work is the proliferation of functional imaging n-back and PASAT studies in the last decade, which allow for more focused meta-analyses. Owen (2005) examined patterns of activation in the n-back, but included both PET and fMRI data in their analyses, as well as a wide variety of n-back difficulty levels, stimuli, and control tasks for fMRI contrasts. On a similar note, Rottschy (2012) examined working memory using a heterogeneous sample of tasks, conditions, and contrasts. While these studies were aimed at broad conceptual differences among working memory paradigms in the literature, our study was more restrictive in its inclusion criteria to focus on specific tasks.

One limitation in noted by Shackman (2011), is that individual differences in macroscopic anatomy, of the cingulate particularly, make it difficult to draw conclusions about the region's functional organization. This limitation may be mitigated by modeling regional activity around a center-of-mass coordinate (i.e. ALE) but serves as a reminder that the spatial resolution in ALE analyses is limited by input data, parameters, thresholds, and other factors. The reporting of peak, rather than center-of-mass, coordinates in fMRI experiments also represents a methodological limitation of ALE. This is further compounded by the tendency to report only one peak, or a small number of peaks for extremely large clusters of activation. To some extent this limitation is inherent to ALE methodology, but could be mitigated by further developing standardized rules for reporting fMRI contrast data.

There are two additional limitations of these meta-analyses that may have impacted results. First, sample size for the 2-back was twice as great as the 3-back and PASAT. This is partially controlled for in the individual ALEs through the use of cluster-level family-wise error, however it still may have an impact on the relative power for identifying clusters. Second, there are more men included in the studies of the 2-back, with an even split for the 3-back and PASAT. Although unlikely, it is possible that this sex imbalance may have impacted the results to some extent.

*4.7. Conclusions and Future Directions*

These meta-analyses include the first ever ALE meta-analyses of PASAT, indicating that it may engage core verbal working memory networks to a greater degree than the n-back, while the 3-back appears to engage more spatial attention. Individual ALE analyses suggest involvement of emotional processing and salience network regions (insula, cingulate) in addition to the well-established verbal working memory regions (Broca's region, bilateral SMA, premotor, posterior parietal cortices) in all 3 tasks. Further, these data illustrate the sensitivity of ALE meta-analysis to discern differences in activation across verbal working memory studies that may be associated with specific cognitive and emotional aspects of these tasks. Further work could identify the extent to which emotion-related regions are activated during these tasks and what factors influence this activity. Further parametric work examining these tasks is necessary to determine more precisely the causes of the activation patterns revealed here.

**Funding**
This study received no external funding. Data were collected from web-accessible journal articles.

**Abbreviations used in this manuscript:**

Terminology:
VWM = Verbal Working Memory
PASAT = Paced Auditory Serial Addition Test
ALE = Activation Likelihood Estimation
MS = Multiple Sclerosis
TBI = Traumatic Brain Injury
PANAS = Positive and Negative Affect Schedule
PRISMA = Preferred Reporting Items for Systematic Reviews and Meta-Analyses

Brain Regions:
BA = Brodmann's Area
SMA = supplementary motor area
DLPFC = dorsolateral prefrontal cortex
aMCC = anterior mid-cingulate cortex
FEF = frontal eye fields
medFG = medial frontal gyrus
SFG = superior frontal gyrus
MFG = middle frontal gyrus
IFG = inferior frontal gyrus

An ALE meta-analytic comparison of verbal working memory tasks                                           27Gronwall, D., & Wrightson, P. (1974). Delayed recovery of intellectual function after minor head injury. *The Lancet*, *304*(7881), 605-609.

Jaeggi, S. M., Buschkuehl, M., Perrig, W. J., & Meier, B. (2010). The concurrent validity of the N-back task as a working memory measure. *Memory*, *18*(4), 394-412.

Laird, A.R., Fox, P.M., Price, C.J., Glahn, D.C., Uecker, A.M., Lancaster, J.L., Turkeltaub, P.E., Kochunov, P., Fox, P.T., 2005. ALE meta - analysis: controlling the false discovery rate and performing statistical contrasts. Human. Brain Mapp. 25, 155 – 164.

LeDoux, J. (2003). The emotional brain, fear, and the amygdala. *Cellular and molecular neurobiology*, *23*(4), 727-738.

Leyro T. M., Zvolensky M. J., & Bernstein A. (2010). Distress tolerance and psychopathological symptoms and disorders: a review of the empirical literature among adults. *Psychological bulletin*, *136*(4), 576.

Mainero, C., Caramia, F., Pozzilli, C., Pisani, A., Pestalozza, I., Borriello, G., ... & Pantano, P. (2004). fMRI evidence of brain reorganization during attention and memory tasks in multiple sclerosis. *Neuroimage*, *21*(3), 858-867.

Moher, D., Liberati, A., Tetzlaff, J., Altman, D. G., & Prisma Group. (2009). Preferred reporting items for systematic reviews and meta-analyses: the PRISMA statement. *PLoS med*, *6*(7), e1000097.

Owen, A. M., McMillan, K. M., Laird, A. R., & Bullmore, E. (2005). N-back working memory paradigm: A meta-analysis of normative functional neuroimaging studies. *Human brain mapping*, *25*(1), 46-59.

Parmenter, B. A., Shucard, J. L., Benedict, R. H., & Shucard, D. W. (2006). Working memory deficits in multiple sclerosis: Comparison between the n-back task and the Paced Auditory Serial Addition Test. *Journal of the International Neuropsychological Society*, *12*(05), 677-687.

Phan, K. L., Wager, T., Taylor, S. F., & Liberzon, I. (2002). Functional neuroanatomy of emotion: a meta-analysis of emotion activation studies in PET and fMRI. *Neuroimage*, *16*(2), 331-348.

Rao, S. M. (1986). Neuropsychology of multiple sclerosis: a critical review. *Journal of Clinical and experimental Neuropsychology*, *8*(5), 503-542.

Rao, V., & Lyketsos, C. (2000). Neuropsychiatric sequelae of traumatic brain injury. *Psychosomatics*, *41*(2), 95-103.

Rottschy, C., Langner, R., Dogan, I., Reetz, K., Laird, A. R., Schulz, J. B., ... & Eickhoff, S. B. (2012). Modelling neural correlates of working memory: a coordinate-based meta-analysis. *Neuroimage*, *60*(1), 830-846.

Sweet, L. H., Rao, S. M., Primeau, M., Mayer, A. R., & Cohen, R. A. (2004). Functional magnetic resonance imaging of working memory among multiple sclerosis patients. *Journal of Neuroimaging*, *14*(2), 150-157.

Tombaugh, T. N. (2006). A comprehensive review of the paced auditory serial addition test (PASAT). *Archives of clinical neuropsychology*, *21*(1), 53-76.

Turkeltaub, P.E., Eden, G.F., Jones, K.M., Zeffiro, T.A., 2002. Meta-analysis of the functional neuroanatomy of single-word reading: method and validation. NeuroImage 16, 765– 780.

Table 1 (continued on next page). Studies included in each meta-analysis. HC = healthy controls, TAL = Talairach coordinates, MNI = Montreal Neurological Institute coordinates.

**2-back**

| Author | year | N | group | task | stimulus | contrast | space |
|---|---|---|---|---|---|---|---|
| Barch | 2007 | 120 | HC | 2back | words | 2back – encoding | TAL |
| Bartova | 2015 | 42 | HC | 2back | numbers | 2back – 0back | TAL |
| Bertolino | 2010 | 28 | HC | 2back | numbers | 2back – 0back | TAL |
| Bleich-Cohen | 2016 | 20 | HC | 2back | numbers | 2back – 0back | TAL |
| Carlson | 1998 | 7 | HC | 2back | shapes | 2back – 0back | TAL |
| Cerasa | 2008 | 30 | HC | 2back | shapes | 2back – 0back | TAL |
| Chang | 2010 | 21 | HC | 2back | letters | 2back – rest | TAL |
| Cousijn | 2010 | 41 | HC | 2back | numbers | 2back – 0back | MNI |
| Deckersbach | 2008 | 17 | HC | 2back | letters | 2back – fixation | MNI |
| Drobyshevsky | 2006 | 31 | HC | 2back | letters | 2back – 0back | TAL |
| Duggirala | 2016 | 50 | HC | 2back | words | 2back – 0back | MNI |
| Fernández-Corcuera | 2013 | 41 | HC | 2back | letters | 2back – baseline | MNI |
| Garrett | 2011 | 19 | HC | 2back | letters | 2back – 0back | MNI |
| Gillis | 2016 | 15 | HC | 2back | letters | 2back – 0back | TAL |
| Habel | 2007 | 21 | HC | 2back | letters | 2back – 0back | MNI |
| Harding | 2016 | 25 | HC | 2back | numbers | 2back – 0back | MNI |
| Honey | 2002 | 20 | HC | 2back | letters | 2back – 0back | TAL |
| Johannsen | 2013 | 12 | HC | 2back | letters | 2back – 0back | MNI |
| Ko | 2013 | 20 | HC | 2back | letters | 2back – 0back | TAL |
| Koppelstaetter | 2008 | 15 | HC | 2back | letters | 2back – 0back | TAL |
| Matsuo | 2007 | 15 | HC | 2back | numbers | 2back – 0back | TAL |
| Meisenzahl | 2006 | 12 | HC | 2back | letters | 2back – 0back | TAL |
| Migo | 2015 | 11 | HC | 2back | letters | 2back – 0back | MNI |
| Oflaz | 2014 | 9 | HC | 2back | letters | 2back – 0back | MNI |
| Park | 2011 | 10 | HC | 2back | letters | 2back – 0back | MNI |
| Quidé | 2013 | 28 | HC | 2back | letters | 2back – 0back | MNI |
| Rodríguez-Cano | 2014 | 52 | HC | 2back | letters | 2back – 0back | MNI |
| Scheuerecker | 2008 | 23 | HC | 2back | letters | 2back – 0back | MNI |
| Seo | 2014 | 34 | HC | 2back | letters | 2back – 0back | MNI |
| Seo | 2012 | 22 | HC | 2back | letters | 2back – 0back | MNI |
| Stretton | 2012 | 15 | HC | 2back | shapes | 2back – 0back | MNI |
| Walitt | 2016 | 13 | HC | 2back | letters | 2back – 0back | MNI |
| Schmidt | 2009 | 25 | HC | 2back | letters | 2back – 0back | TAL |
| Schmidt | 2009 | 21 | HC | 2back | letters | 2back – 0back | TAL |

### 3-back

| Author | year | N | group | task | stimulus | contrast | space |
|---|---|---|---|---|---|---|---|
| D'Esposito | 1998 | 24 | HC | 3back | letters | 3back – 0back | TAL |
| Honey | 2002 | 10 | HC | 3back | letters | 3back – 0back | TAL |
| Mizuno | 2008 | 14 | HC | 3back | numbers | 3back – 0back | TAL |
| Ravizza (1.5T-scan) | 2004 | 10 | HC | 3back | letters | 3back – 0back | TAL |
| Ravizza (3T-scan) | 2004 | 11 | HC | 3back | letters | 3back – 0back | TAL |
| Belayachi | 2015 | 18 | HC | 3back | letters | 3back – 0back | TAL |
| Drapier | 2008 | 20 | HC | 3back | letters | 3back – 0back | TAL |
| Elbin | 2012 | 14 | HC | 3back | letters | 3back – 0back | MNI |
| Gaudeau-Bosma | 2013 | 19 | HC | 3back | letters | 3back – 0back | MNI |
| Haller | 2005 | 16 | HC | 3back | letters | 3back – 0back | TAL |
| Landré | 2012 | 16 | HC | 3back | letters | 3back – control | MNI |
| Surguladze | 2007 | 8 | HC | 3back | letters | 3back – 0back | TAL |
| van Ruitenbeek | 2013 | 16 | HC | 3back | letters | 3back – 0back | MNI |
| Gunderson | 2008 | 13 | HC | 3back | letters | 3back – fixation | TAL |
| Cader | 2006 | 16 | HC | 3back | letters | 3back – 1back | MNI |
| Martinkauppi | 2000 | 10 | HC | 3back | letters | 3back – 1back | TAL |
| Thaler | 2016 | 39 | HC | 3back | letters | 3back – 0back | TAL |

### PASAT

| Author | year | N | group | task | stimulus | contrast | space |
|---|---|---|---|---|---|---|---|
| Audoin | 2005 | 10 | HC | PASAT | auditory | PASAT – control task | TAL |
| Audoin | 2005 | 18 | HC | PASAT | auditory | PASAT – control task | TAL |
| Audoin | 2003 | 10 | HC | PASAT | auditory | PASAT – control task | TAL |
| Bonzano | 2009 | 18 | HC | PVSAT | visual | PVSAT – control task | TAL |
| Cardinal | 2008 | 10 | HC | PASAT | auditory | PASAT – control task | TAL |
| Christodoulou | 2001 | 7 | HC | modified PASAT | auditory | mPASAT – imagination | TAL |
| Forn | 2012 | 15 | HC | PASAT | auditory | PASAT – control task | MNI |
| Forn | 2011 | 17 | HC | PASAT | auditory | PASAT – control task | TAL |
| Forn | 2008 | 13 | HC | covert PASAT | auditory | cPASAT – control task | TAL |
| Forn | 2006 | 10 | HC | covert PASAT | auditory | cPASAT – control task | TAL |
| Hayter | 2007 | 15 | HC | PASAT | auditory | PASAT – control task | MNI |
| Koric | 2012 | 15 | HC | PASAT | auditory | PASAT – control task | TAL |
| Lazeron | 2003 | 9 | HC | PVSAT | visual | cPVSAT – number recall | TAL |
| Mainero | 2004 | 22 | HC | modified PASAT | auditory | PASAT – recall task | TAL |
| Maruishi | 2007 | 12 | HC | PVSAT | visual | PVSAT – control task | MNI |
| Tudos | 2014 | 20 | HC | PASAT+PVSAT | conjunction | PASAT/PVSAT – control task | MNI |

Table 2. ALE Clusters identified for task contrasts

| Clusters | Volume (mm³) | X | Y | Z |
|---|---|---|---|---|
| 3-back > 2-back | | | | |
| 1  L/R Cingulate, R medFG, R SFG | 1488 | 5 | 20.1 | 42.8 |
| | | | | |
| 3-back > PASAT | | | | |
| 1  R MFG | 232 | 35.7 | 28.2 | 30.6 |
| 2  R SFG | 200 | 5.4 | 20.3 | 50.7 |
| | | | | |
| PASAT > 2-back | | | | |
| 1  L/R medFG, L/R SFG | 664 | -1.8 | 1.6 | 60.6 |
| 2  R Precentral Gyrus, R IFG | 488 | 50.1 | 4.1 | 28.6 |
| 3  L Supramarginal Gyrus | 256 | -36.4 | -43.7 | 44.9 |
| 4  L Superior Parietal Lobule | 240 | -31.6 | -55.5 | 52.6 |
| | | | | |
| PASAT > 3-back | | | | |
| 1  L/R medFG, L SFG | 912 | -4.3 | 0.5 | 59.7 |
| 2  L MFG | 392 | -43.4 | 30.6 | 23 |
| 3  L Supramarginal Gyrus | 320 | -35.7 | -44.1 | 37.8 |
| 4  L Superior Parietal Lobule | 208 | -28.4 | -57.5 | 49.6 |

Table 3. Identified local maxima within ALE clusters

| Clusters | | X | Y | Z |
|---|---|---|---|---|
| 3-back > 2-back | | | | |
| 1 | R SFG | 6 | 25 | 42 |
| | R Cingulate | 4 | 18 | 48 |
| | R Cingulate | 4 | 21 | 34 |
| | | | | |
| 3-back > PASAT | | | | |
| 1 | R MFG | 34 | 28 | 30 |
| | R MFG | 42 | 24 | 32 |
| 2 | R SFG | 4 | 22 | 50 |
| | | | | |
| PASAT > 2-back | | | | |
| 1 | L medFG | -4 | 4 | 62 |
| 2 | R Precentral Gyrus | 51 | 3 | 29 |
| 3 | L Inferior Parietal Lobule | -34 | -46 | 48 |
| | L Inferior Parietal Lobule | -38 | -40 | 42 |
| 4 | L Precuneus | -28 | -54 | 50 |
| | L Superior Parietal Lobule | -32 | -52 | 52 |
| | | | | |
| PASAT > 3-back | | | | |
| 1 | L medFG | -6 | 2 | 62 |
| | L medFG | -10 | 0 | 58 |
| | L medFG | -10 | 8 | 52 |
| 2 | L MFG | -44 | 30 | 20 |
| | L MFG | -46 | 32 | 24 |
| 3 | L Inferior Parietal Lobule | -34 | -44 | 38 |
| 4 | L Precuneus | -26 | -58 | 50 |

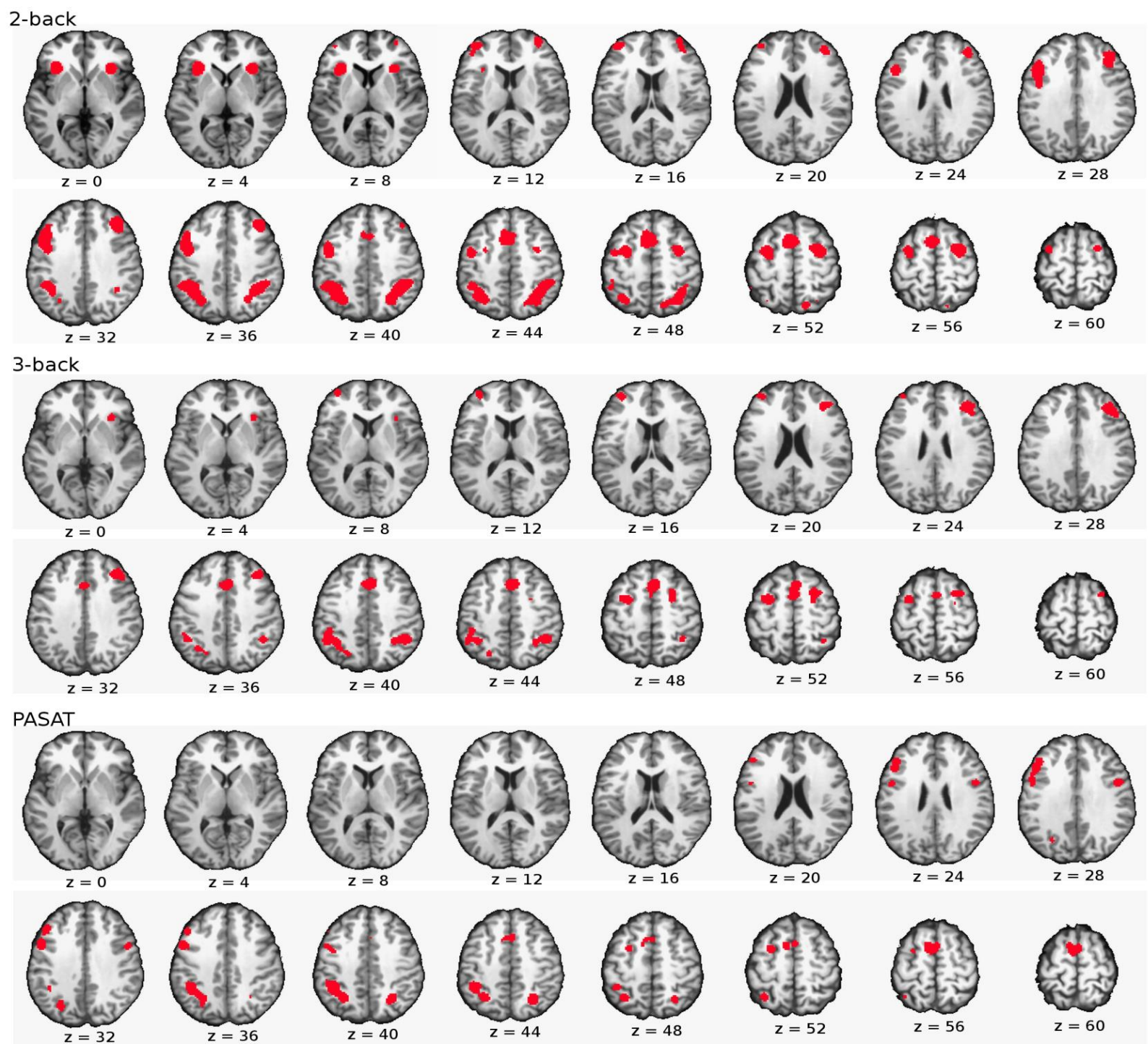

Figure 1. Meta-analytic activation maps for the 2-back (top), 3-back (middle), and PASAT (bottom). Regions that meet the cluster-level family-wise error threshold ($< 0.05$) are depicted in red. Talairach z-coordinates are listed for each axial slice.

## 3-back > 2-back

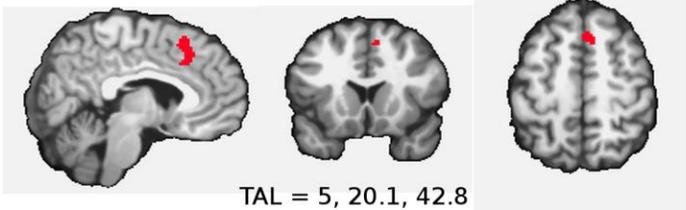
TAL = 5, 20.1, 42.8

## 3-back > PASAT

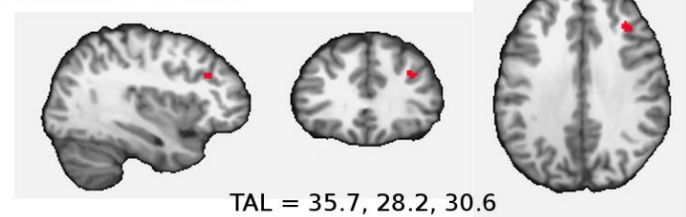
TAL = 35.7, 28.2, 30.6

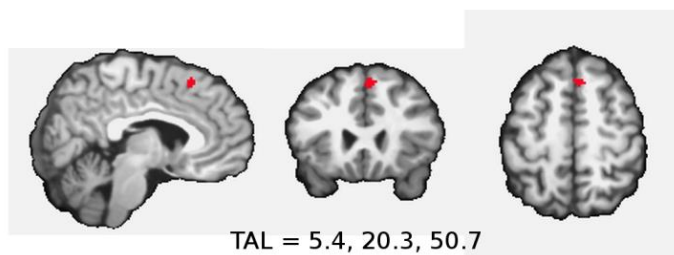
TAL = 5.4, 20.3, 50.7

## PASAT > 2-back

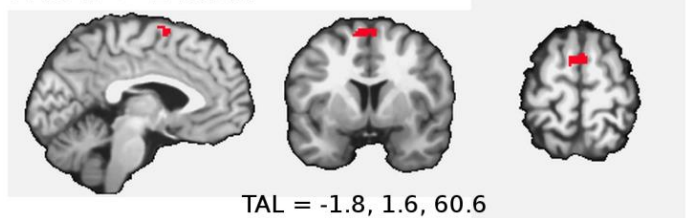
TAL = -1.8, 1.6, 60.6

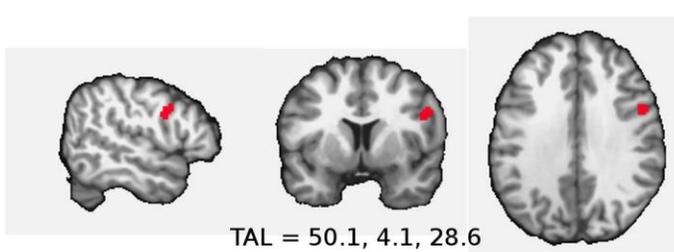
TAL = 50.1, 4.1, 28.6

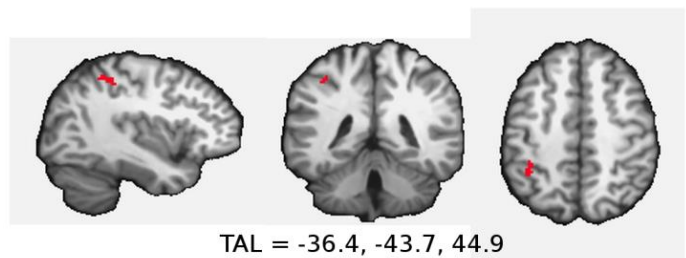
TAL = -36.4, -43.7, 44.9

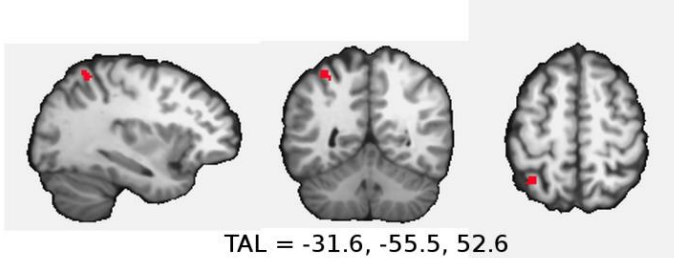
TAL = -31.6, -55.5, 52.6

## PASAT > 3-back

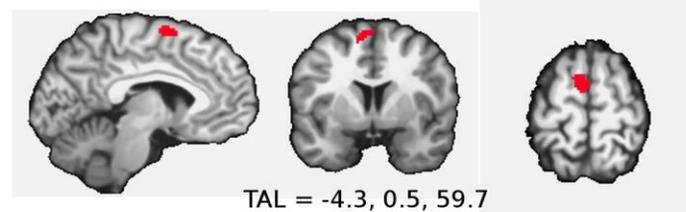
TAL = -4.3, 0.5, 59.7

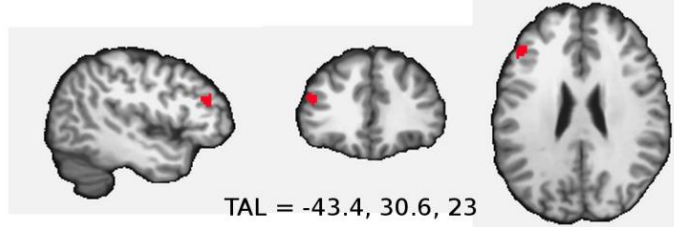
TAL = -43.4, 30.6, 23

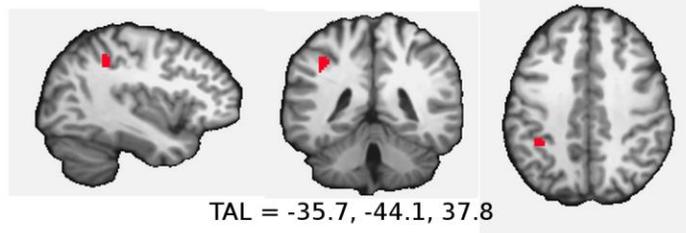
TAL = -35.7, -44.1, 37.8

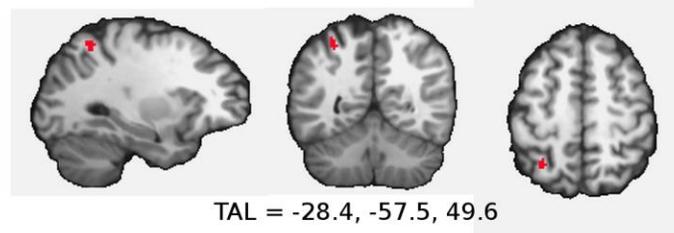
TAL = -28.4, -57.5, 49.6

Figure 2. Sagittal, coronal, and axial view of each cluster identified by meta-analytic comparison of the 2-back, 3-back, and PASAT. Center-of-mass coordinates in Talairach space are listed for each cluster.

**Supplementary Materials**

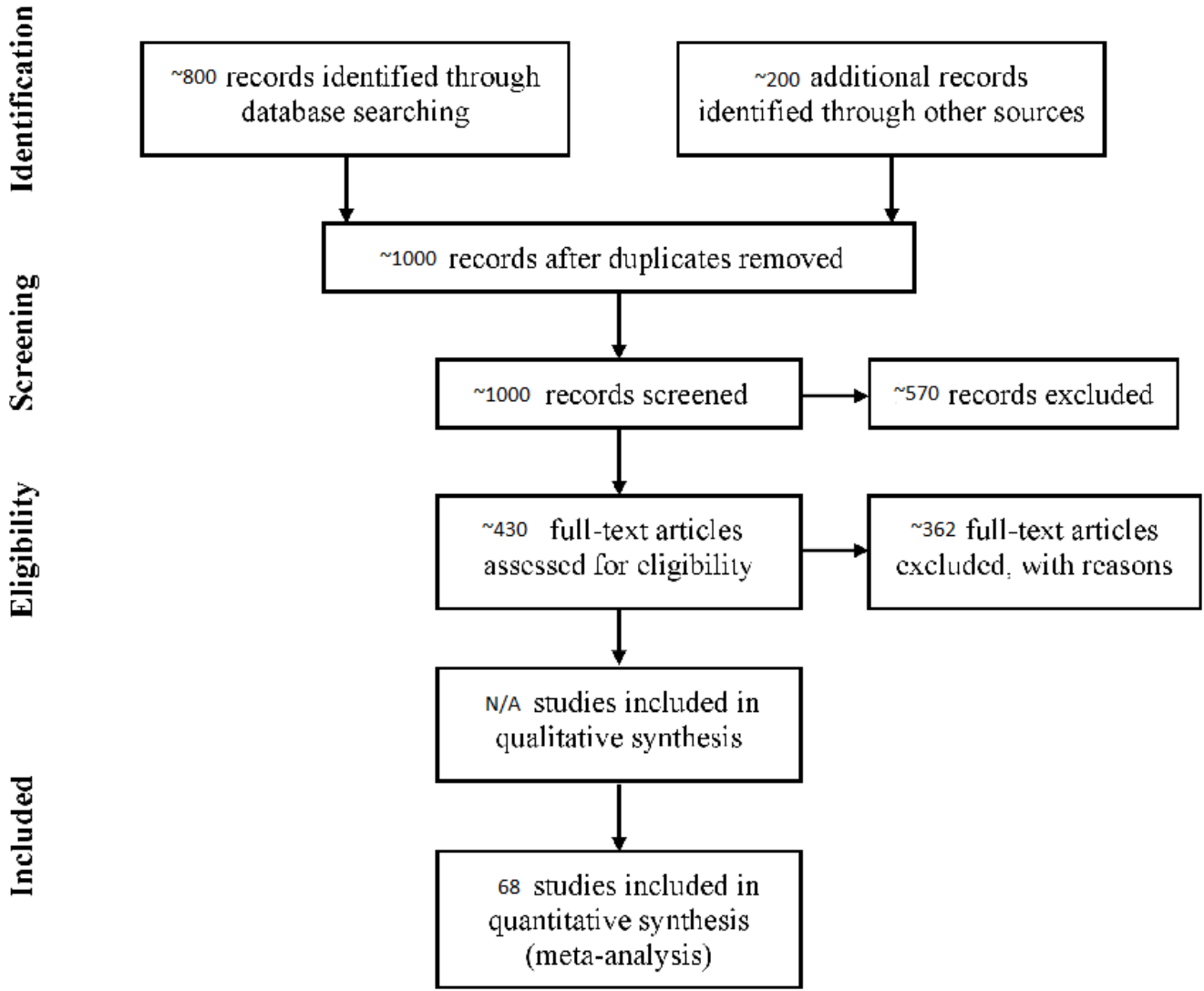

Supplementary Figure A. Description of screening process used for selection of studies (PRISMA).

Supplementary Table 1. (continued on next page)  Demographics information and task performance for the 2-back, 3-back, and PASAT.  NR denotes information that was not reported in the paper.

| Author | year | N | N (men) | N (women) | Average Age | Response modality | Task performance (%correct) |
| --- | --- | --- | --- | --- | --- | --- | --- |
| Barch | 2007 | 120 | 50 | 70 | 27.2 | Right hand | 95 |
| Bartova | 2015 | 42 | 17 | 25 | 25.3 | Right hand | 81 |
| Bertolino | 2010 | 28 | NR | NR | 23.5 | NR | NR |
| Bleich-Cohen | 2016 | 20 | 12 | 8 | 26.4 | Right hand | 99.38 |
| Carlson | 1998 | 7 | 4 | 3 | 21.1 | Right hand | 96.6 |
| Cerasa | 2008 | 30 | 30 | 0 | 30.5 | Right hand | 77.75 |
| Chang | 2010 | 21 | 21 | 0 | 49.7 | Right hand | 70.1 |
| Cousijn | 2010 | 41 | 41 | 0 | 26.5 | Right hand | 85 |
| Deckersbach | 2008 | 17 | 0 | 17 | 25.6 | Right hand | 94.43 |
| Drobyshevsky | 2006 | 31 | 16 | 15 | 40.9 | Right hand | 90.4 |
| Duggirala | 2016 | 50 | 28 | 22 | 23.62 | Right hand | 79.77 |
| Fernández-Corcuera | 2013 | 41 | 27 | 17 | 40.27 | Right hand | NR |
| Garrett | 2011 | 19 | 13 | 6 | 34.85 | Right hand | 99.77 |
| Gillis | 2016 | 15 | 15 | 0 | 25.13 | Right hand | 95.83 |
| Habel | 2007 | 21 | 21 | 0 | 30.77 | Right hand | 94 |
| Harding | 2016 | 25 | 14 | 11 | 25.5 | Right hand | 94.9 |
| Honey | 2002 | 20 | 20 | 0 | 39.3 | Right hand | 81.92 |
| Johannsen | 2013 | 12 | 4 | 8 | 26.1 | Right hand | 73.8 |
| Ko | 2013 | 20 | 20 | 0 | 26.3 | Right hand | 84.05 |
| Koppelstaetter | 2008 | 15 | 15 | 0 | 36 | Right hand | 95.24 |
| Matsuo | 2007 | 15 | 6 | 9 | 37.7 | Dominant hand | 71.5 |
| Meisenzahl | 2006 | 12 | 11 | 1 | 33.58 | Right hand | 97.57 |
| Migo | 2015 | 11 | 7 | 4 | 70.27 | NR | 99 |
| Oflaz | 2014 | 9 | 7 | 2 | 44.6 | Right hand | 82.75 |
| Park | 2011 | 10 | 10 | 0 | 23.7 | Right hand | NR |
| Quidé | 2013 | 28 | 14 | 14 | 32.96 | NR | 75 |
| Rodríguez-Cano | 2014 | 52 | 20 | 32 | 46.25 | Right hand | NR |
| Scheuerecker | 2008 | 23 | 19 | 4 | 32.6 | Right hand | 99 |
| Seo | 2014 | 34 | 0 | 34 | 59.3 | Right hand | 77.2 |
| Seo | 2012 | 22 | 0 | 22 | 38.27 | Right hand | 95.56 |
| Stretton | 2012 | 15 | 4 | 11 | 27 | Dominant hand | 68.1 |
| Walitt | 2016 | 13 | 0 | 13 | 44.2 | Right hand | 64.7 |
| Schmidt | 2009 | 25 | 25 | 0 | 34.36 | Right hand | 83.2 |
| Schmidt | 2009 | 21 | 0 | 21 | 33.13 | Right hand | 93.95 |

**2-back**

**3-back**

| Author | year | N | N (men) | N (women) | Average Age | Response modality | Task performance (%correct) |
|---|---|---|---|---|---|---|---|
| D'Esposito | 1998 | 16 | 10 | 6 | 23.8 | Right hand | 87.6 |
| Honey | 2002 | 10 | 7 | 3 | 25.7 | Right hand | 80 |
| Mizuno | 2008 | 14 | 7 | 7 | 22.4 | Right hand | 80.87 |
| Ravizza (1.5T-scan) | 2004 | 10 | NR | NR | 27.5 | Right hand | 92 |
| Ravizza (3T-scan) | 2004 | 11 | NR | NR | 27.5 | Right hand | 92 |
| Belayachi | 2015 | 18 | 8 | 10 | 23.5 | Right hand | 65 |
| Drapier | 2008 | 20 | 10 | 10 | 41.9 | NR | 79.46 |
| Elbin | 2012 | 14 | NR | NR | NR | NR | 76.66 |
| Gaudeau-Bosma | 2013 | 19 | 11 | 8 | 31.6 | Right hand | 78.7 |
| Haller | 2005 | 16 | 8 | 8 | 25.2 | Right hand | 83.6 |
| Landré | 2012 | 16 | 0 | 16 | 24.8 | Dominant hand | NR |
| Surguladze | 2007 | 8 | 4 | 4 | 32.8 | NR | 69.33 |
| van Ruitenbeek | 2013 | 16 | 8 | 8 | 24.5 | Right hand | 69.09 |
| Gunderson | 2008 | 13 | 13 | 0 | 28 | Right hand | 84 |
| Cader | 2006 | 16 | 6 | 10 | 39 | Right hand | 72 |
| Martinkauppi | 2000 | 10 | 5 | 5 | 25 | Right hand | 76 |
| Thaler | 2016 | 39 | 19 | 20 | 46.33 | NR | 59 |

**PASAT**

| Author | year | N | N (men) | N (women) | Average Age | Response modality | Task performance (%correct) |
|---|---|---|---|---|---|---|---|
| Audoin | 2005 | 18 | 5 | 13 | 25.3 | Vocal | 81.2 |
| Audoin | 2003 | 10 | 8 | 2 | 26.1 | Vocal | 78 |
| Bonzano | 2009 | 18 | 9 | 9 | 32.5 | Vocal | 90.58 |
| Cardinal | 2008 | 10 | 3 | 7 | 34.4 | Vocal | 95* |
| Christodoulou | 2001 | 7 | 4 | 3 | 29.71 | Right hand | 94.05** |
| Forn | 2012 | 15 | NR | NR | 32.3 | Vocal | 87.5 |
| Forn | 2011 | 17 | 7 | 10 | 32.76 | Vocal | 78.25 |
| Forn | 2008 | 13 | 8 | 5 | 21.9 | Vocal | 87.37 |
| Forn | 2006 | 10 | 5 | 5 | 32.73 | Silent repetition | 65.6 |
| Hayter | 2007 | 15 | 6 | 9 | 23.5 | Vocal | 82 |
| Koric | 2012 | 15 | 6 | 9 | 34.7 | Vocal | 84.62 |
| Lazeron | 2003 | 9 | 6 | 3 | 24 | Mental summation | NR |
| Mainero | 2004 | 22 | 11 | 11 | NR | Right hand | 78.8** |
| Maruishi | 2007 | 12 | 11 | 1 | 26.4 | Right hand | 100** |
| Tudos | 2014 | 20 | 10 | 10 | 23 | Right hand | 95.9** |

* : Performance out of scanner, after practice.
**: Task asks subjects to determine if two prior numbers sum to a specific number (i.e.10,8).